\documentclass[journal]{IEEEtran}
\ifCLASSINFOpdf
  \usepackage[pdftex]{graphicx}
  \DeclareGraphicsExtensions{.pdf,.jpeg,.png,.jpg}
\else
\fi
\usepackage{amsmath}
\usepackage{amssymb}
\begin{document}
%
\title{Dynamics of a magnetic gear\\ with two cogging-free operation modes}
%
%
%

\author{Stefan~Hartung
    and~Ingo~Rehberg
\thanks{S. Hartung and I. Rehberg are with the Department
of Experimental Physics V, University of Bayreuth,
Germany, e-mail: stefan.hartung@uni-bayreuth.de.}
\thanks{Manuscript received July 1, 2020}}

%
%

\markboth{Submitted to IEEE Transactions on Magnetics}%
{Shell \MakeLowercase{\textit{et al.}}: Bare Demo of IEEEtran.cls for IEEE Journals}
%



\maketitle

\begin{abstract}
The coupling of two rotating spherical magnets is investigated experimentally. For two specific angles between the input and output rotation axes, a cogging-free coupling is observed, where the driven magnet is phase-locked to the driving one. The striking difference between these two modes of operation is the reversed sense of rotation of the driven magnet. For other angles the experiments reveal a more complex dynamical behavior, which is divided in three different classes. This is done by analysing the deviation from a periodic motion of the driven magnet, and by measuring the total harmonic distortion of this rotation. The experimental results can be understood by a mathematical model based on pure dipole-dipole interaction, with the addition of adequate friction terms.
\end{abstract}


%
\IEEEpeerreviewmaketitle

\section{Introduction}
\IEEEPARstart{M}{agnetic} gears have advantages: The input and output are free of mechanical contact. Thus they are not subject to mechanical wear, need no lubrication, and operate with reduced maintenance.  Moreover, they possess inherent overload protection, are noiseless, and highly reliable \cite{Furlani2001}.

With the appearance of strong magnets based on alloys of rare-earth elements, the interest in magnetic gears based on permanent magnets grew because of increased torque transmission capabilities \cite{Tsurumoto1987}, \cite{wang2009development}, \cite{jian2009magnetic}, \cite{atallah2001novel}, and continues to do so today \cite{park2020comparison}.

An interesting type of a magnetic gear based on pure magnetic dipoles has been proposed in 2015 by J.\ Sch\"onke \cite{Schoenke2015}. Inspired by his former work on a seven-fold magnetic clutch \cite{Schoenke2015a}, he demonstrated theoretically that two magnetic dipoles could couple in two cogging-free modes, provided that the angles of the two rotation axes follow a certain algebraic condition. Almost pure dipoles are indeed commercially available in the form of spherical permanent magnets, as has been demonstrated experimentally \cite{Hartung2018}. A first experimental demonstration of the principle of this gear using such spherical magnets concentrated on static aspects \cite{borgers2018}. In this paper, we provide measurements of the \emph {dynamical} behaviour of such a gear. Moreover, we compare these measurements with numerical simulations of the dynamics of two coupled magnetic dipoles. 

\section{Experimental setup}
\begin{figure}[!t]
\centering
\includegraphics[width=\linewidth]{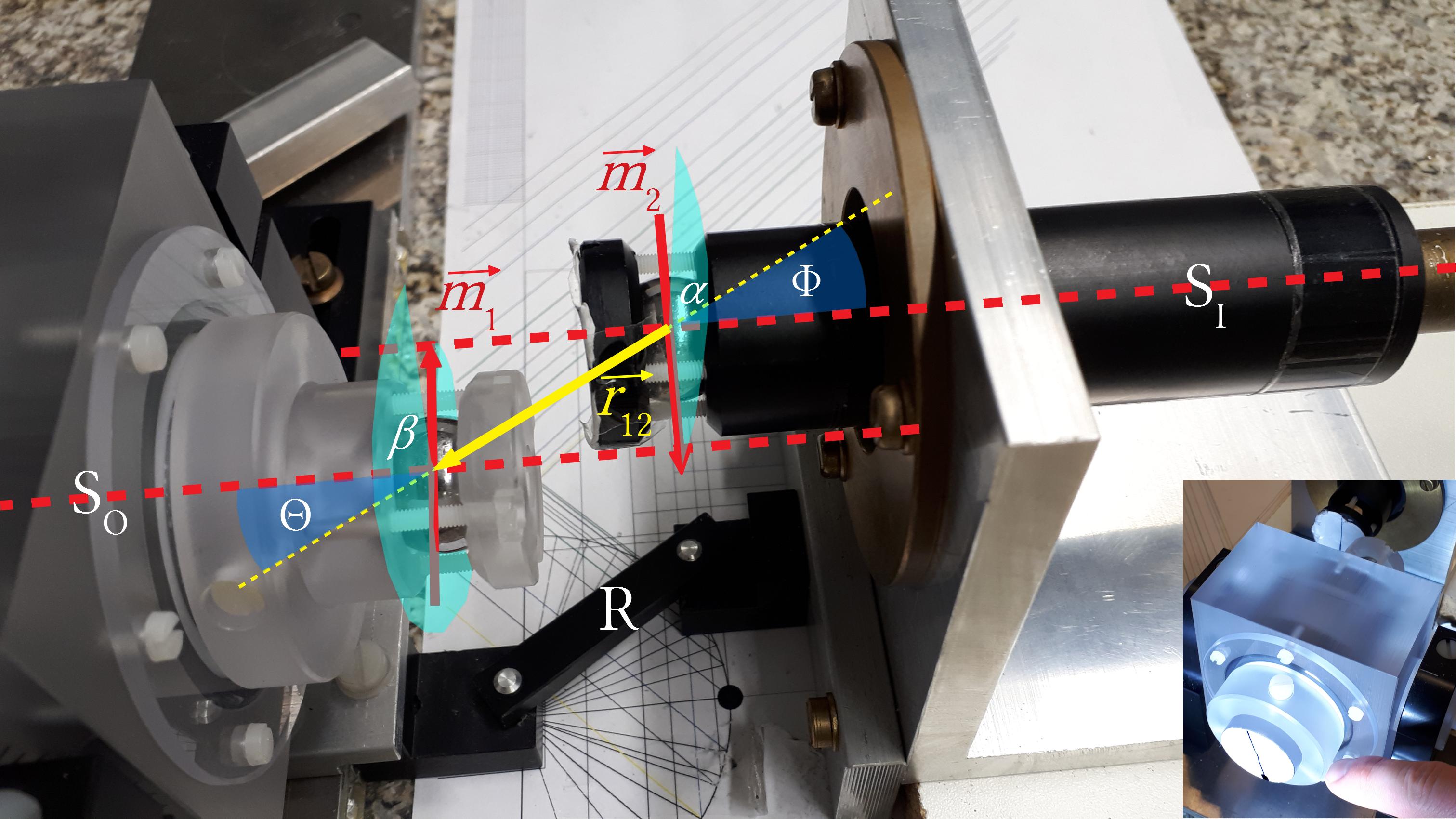}
\caption{Photograph of the experimental setup. Both shaft axes are marked by a dashed red line. The input shaft connected to the motor by a brass rod is marked as $\text{S}_{\text{I}}$, the output shaft as $\text{S}_{\text{O}}$. The directions of the magnetic dipole moments $\vec{m}_1$ and $\vec{m}_2$ are shown by the red arrows. Their respective rotations are marked by the input dipole angle $\alpha$ and the output dipole angle $\beta$. The spacer holding the distance $\vec{r}_{12}$ between the dipole centers is named R. The dotted yellow line connecting the two dipoles forms the input shaft angle $\Theta$ with the input axis and the output shaft angle $\Phi$ with the output axis. The marker for the optical data readout at the front end of $\text{S}_{\text{O}}$ is seen in the inset.}
\label{fig1}
\end{figure}

Fig.\,\ref{fig1} shows the experimental setup, which is very similar to the one described by Borgers et al.\,\cite{borgers2018}. The two spherical neodymium permanent magnets have a diameter of 19\,mm and are each attached to a shaft with their dipole moment aligned perpendicular to the rotation axis with a precision of about $3^{\circ}$. Both axes lie in the horizontal plane. The bearings are non-magnetic and electrically non-conducting. The input shaft runs in a bearing that consists of a plastic cage with glass beads, which has been constructed in our in-house workshop. It is connected to a stepper motor by an 80\,cm long brass rod to suppress any magnetic interference between the motor and the spherical magnets. The output shaft runs in two industrial full ceramic deep groove ball bearings made out of silicon nitride, in which it can rotate freely \cite{kugellager}. 

\begin{figure}[!t]
\centering
\includegraphics[width=\linewidth]{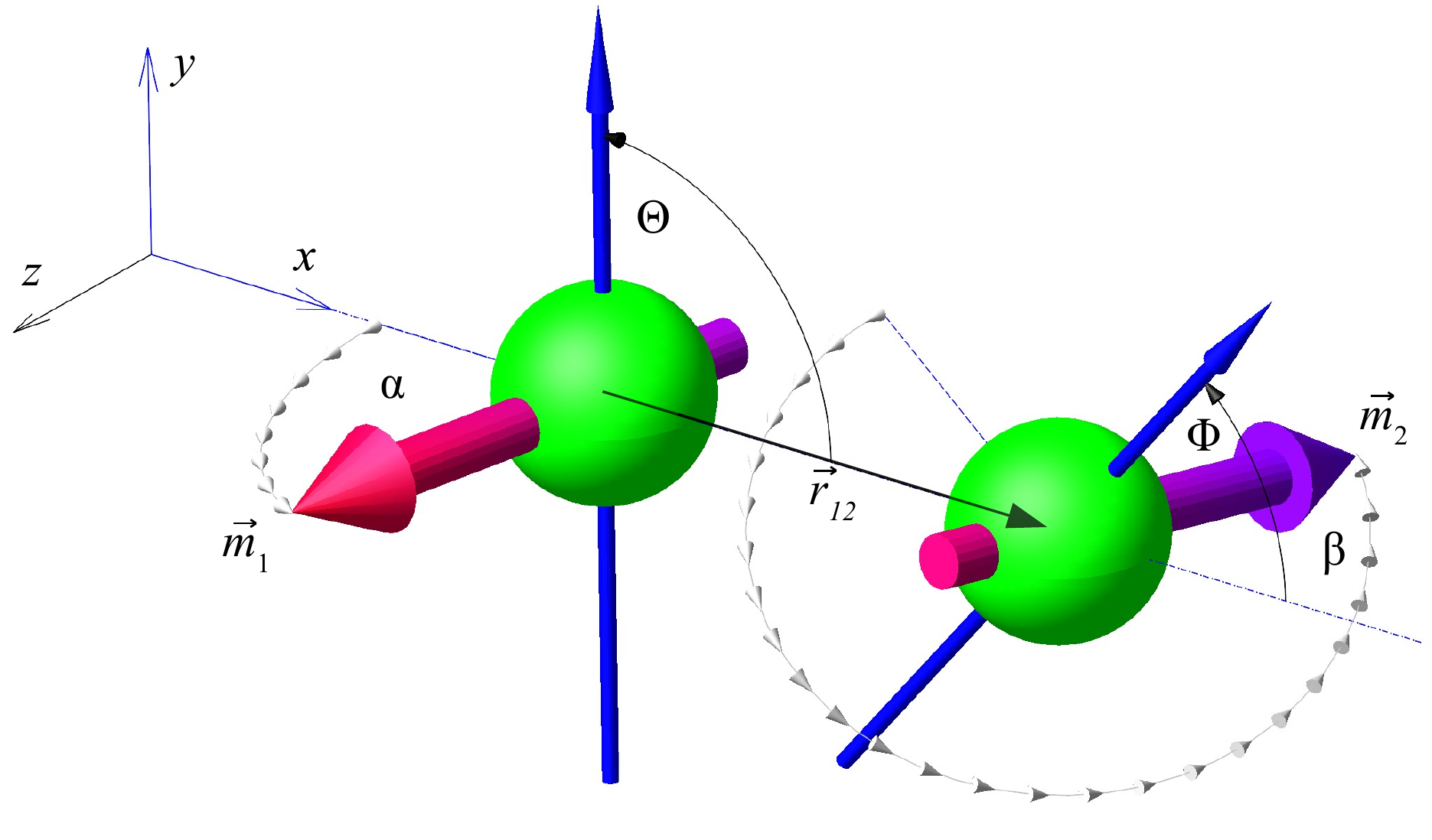}
\caption{Coordinate system showing the relevant angles for the magnet orientation. The $x$-axis connects the center of the input magnet (left sphere) with that of the output magnet (right sphere). Their distance is marked as $\vec{r}_{12}$. The rotation axes of the magnets are shown by the blue arrows and lie in the $x$-$y$-plane. Their angle towards the $x$-axis is $\Theta$ for the input and $\Phi$ for the output. The respective dipole moments $\vec{m}_1$ and $\vec{m}_2$ are shown as thick arrows. Their orientation is marked by the angles $\alpha$ and $\beta$ measured against the $x$-$y$-plane.}
\label{fig2}
\end{figure}

To track the orientation of the dipole $\vec{m}_2$, the end of the output shaft is covered with a white surface marked with a black line which is recorded by a CCD-camera. The stepper motor and the CCD-camera are both connected to a computer. The input dipole angle $\alpha$ (see Fig.\,\ref{fig2}) is detected by a signal from the stepper motor yielding a resolution of $7.2^{\circ}$. The output dipole angle $\beta$ is obtained by digitally processing the image of the black line marker. As an additional feedback to track $\alpha$, a commercial wireless acceleration sensor is located on the brass rod that drives the input \cite{sensor}. The position of the output shaft is fixed on a granite table, while the input shaft can be oriented freely on the surface of the table. An interchangeable spacer keeps the distance between the dipoles constant during a set of experiments while the relative orientation of their rotation axes can be varied. 

The system geometry is further explained in Fig.\,\ref{fig2}. The $x$-axis is determined by the connection between the two dipoles $\vec{r}_{12}$. The $y$-axis lies in the plane defined by $\vec{r}_{12}$ and the shaft axes. The results shown in this article stem from experiments with parallel rotation axes of input and output so that $\Theta$\,$\approx$\,$\Phi$ applies.

\section{Quantifying system parameters}

While the work of Borgers et al. focused on the static equilibrium conditions of a magnetic clutch \cite{borgers2018}, this article shows the dynamic behavior of the output when the input magnet is driven with a constant frequency. Thus, friction is an important additional parameter in our system which we determine experimentally by keeping the input angle fixed and analysing the oscillation of the output around its equilibrium position. Following refs. \cite{Schoenke2015, borgers2018}, the equilibrium position for the angle $\beta_0$ of the output as a function of a fixed input angle $\alpha_0$ can be written as
\begin{equation}
    \beta_0=\mathrm{arctan}\left(\frac{1}{\Delta}\mathrm{tan}(\alpha_0)+k\cdot\pi\right),\,\,\,k \in \mathbb{N}_0.
\label{eq:equilibrium}
\end{equation}
Here $\Delta \neq 0$ is the shaft orientation index of the input and output shaft angles $\Theta$ and $\Phi$ and writes as
\begin{equation}
    \Delta=\cos\Theta \cos\Phi-2\,\sin\Theta \sin\Phi.
\label{eq:Delta}
\end{equation}

In an ideal system without friction, the oscillation the output will undergo when turned out of its equilibrium position is therefore described by
\begin{equation}
    \ddot{\beta}+\kappa\,\sin(\beta-\beta_0)=0,
    \label{eq:nofric}
\end{equation}
where $\kappa$ is the restoring coefficient defined by
\begin{equation}
    \kappa=\left(\frac{d\tau_{\text{ax}}}{d\beta}\right)\cdot\frac{1}{I_{\text{ax}}}=\frac{D}{I_{\text{ax}}}.
    \label{eq:kappa}
\end{equation}
Here $\tau_{\text{ax}}$ is the torque on the output, $I_{\text{ax}}$ is its moment of inertia, both along its rotation axis, and $D$ is the directional constant. For any real coupling, however, this equation is unsatisfactory, since the ball bearings are prone to friction. 

\begin{figure}[!t]
\centering
\includegraphics[width=\linewidth]{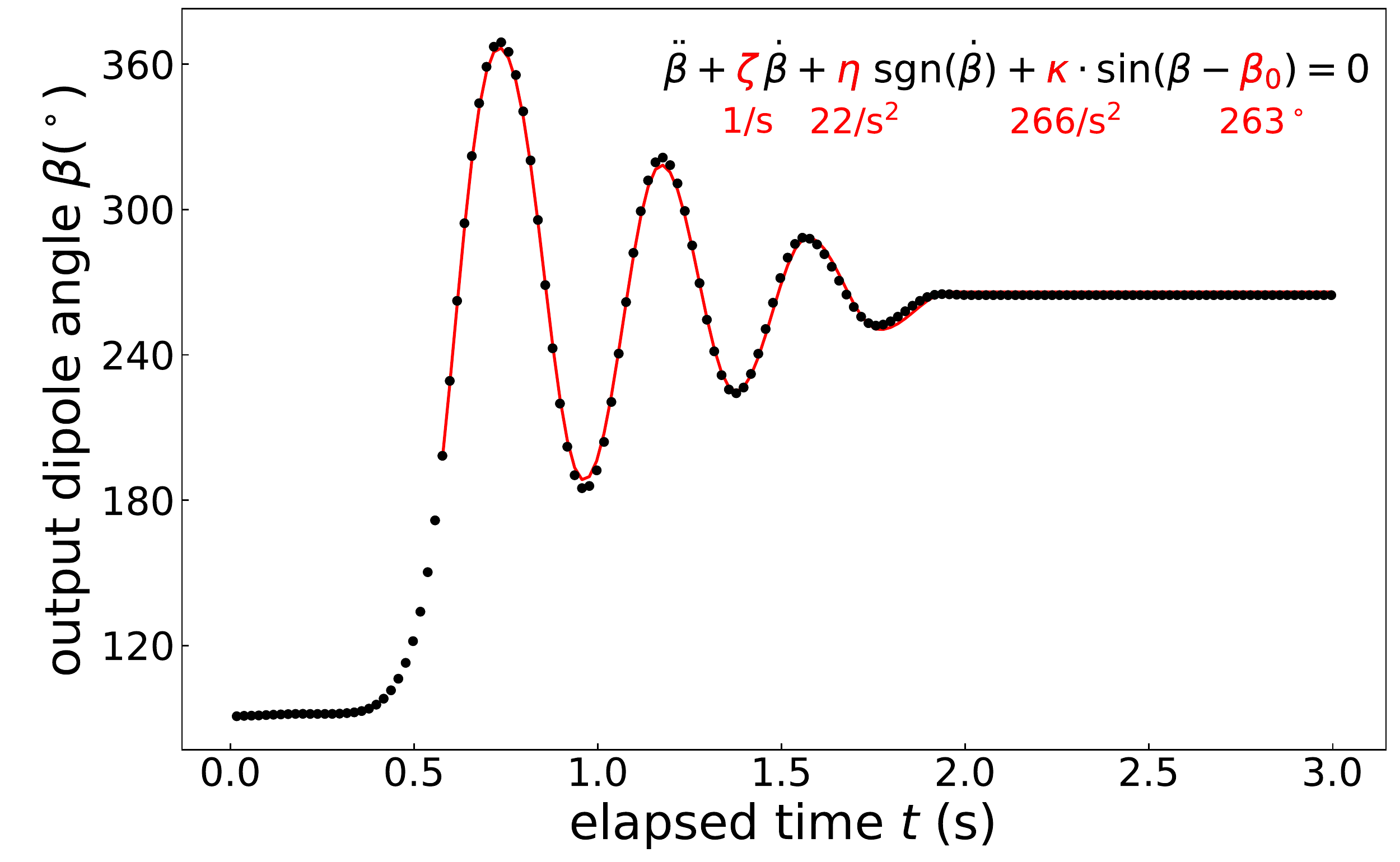}
\caption{Temporal evolution of the output angle $\beta$ for a locked input at angle $\alpha_0$. The parallel shaft angles were held constant at $\Phi$\,=\,$\Theta$\,=\,$31^{\circ}$. The $x$-axis shows the elapsed time after the output was released from the unstable equilibrium position. The measured data points appear as black dots. The red curve depicts the fit to the data according to the formula in the top right where red letters show the fit parameters and their respective values underneath.}
\label{fig3}
\end{figure}

To quantify the influence of friction we conduct an experiment in which the input rotation is prevented while the output magnet can rotate freely. We first turn the output magnet out of its equilibrium position until the restoring moment reaches its maximum. From there we release the output magnet and record its damped oscillation with the camera. An exemplary result of such a measurement can be seen in Fig.\,\ref{fig3}. We make 20 such measurements, each with a different locked input angle $\alpha_0$. The output and input shaft are parallel to each other with constant shaft angles $\Theta$\,=\,$\Phi$\,=\,$(31\pm3)^{\circ}$. The relatively high uncertainty stems from the fact, that a small tilt of the magnets in their sockets is difficult to avoid during their fixation in our setup, as well as small deviations from the parallel alignment.

We evaluate our model by fitting it to the data points. We find that the addition of a single, dry friction-related term is not sufficient to give a good representation of the experiment. The data implies that we need another, rotation frequency depended friction parameter to describe the oscillation. By consideration of these aspects (\ref{eq:nofric}) changes towards
\begin{equation}
    \ddot{\beta}+\zeta\dot{\beta}+\eta\,\mathrm{sgn}(\dot{\beta})+\kappa\,\sin(\beta-\beta_0)=0.
    \label{eq:model}
\end{equation}
Here we call $\zeta$ the damping torque coefficient that takes account for the fluid-like friction in our system. The dry friction is represented by  $\eta$,  the friction torque coefficient, defined by
\begin{equation}
    \eta = \frac{\tau_\text{fr}}{I_{\text{ax}}} = \frac{\mu\,F\,\frac{d_\text{cyl}}{2}}{I_{\text{ax}}},
\label{eq:eta}
\end{equation}
with $\tau_\text{fr}$ being the friction torque for a normal force $F$ on the bearings, a rotating cylinder of diameter $d_\text{cyl}$ and a sliding friction coefficient $\mu$.

\begin{figure}[!t]
\centering
\includegraphics[width=\linewidth]{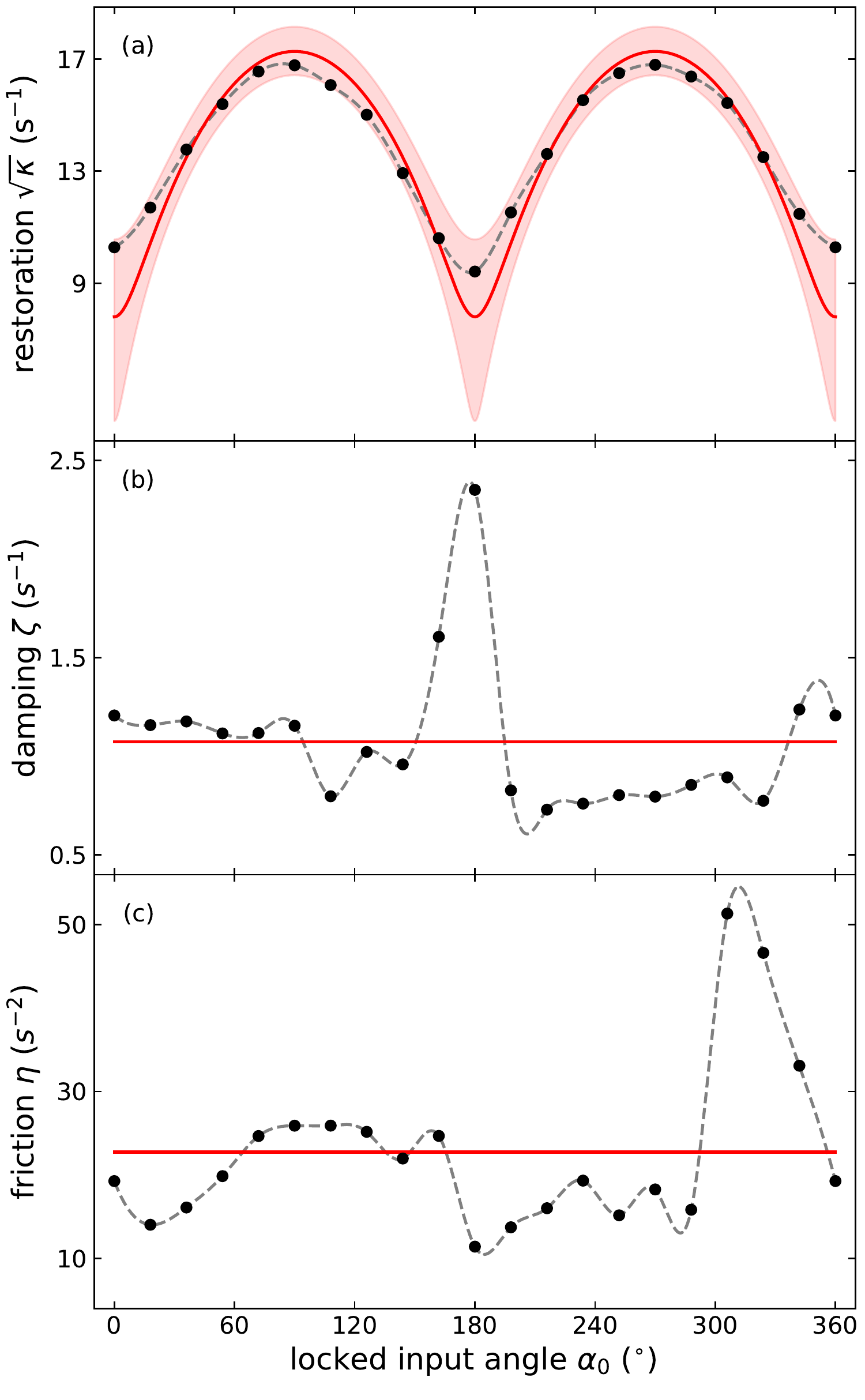}
\caption{Summary of the parameters from fitting (\ref{eq:model}) to the damped oscillation of $\beta$ for different locked input angles $\alpha_0$. The black dots show the obtained parameter values. The grey dotted lines are guides to the eye. In graph (a) the square root of the restoring coefficient $\kappa$ is shown. The red line is the result of a theoretical calculation for $\Theta$\,=\,$31^{\circ}$ with its uncertainty shown by the red area. In (b) the damping  torque  coefficient $\zeta$ can be seen. The red line indicates the arithmetic mean of the data points. Graph (c) shows the friction torque coefficient $\eta$, with the arithmetic mean given by the red line.}
\label{fig4}
\end{figure}

For a realistic model of the output response we need estimates for the values of $\zeta$ and $\eta$. Therefore we fit (\ref{eq:model}) to our sets of data for different $\alpha_0$ with $\zeta$, $\eta$, $\kappa$ as the fit parameters. We decide to fit $\beta_0$ as well. This is more precise than calculating $\beta_0$ from (\ref{eq:equilibrium}). While the value of $\beta_0$ is not interesting, the obtained values of the other fit parameters are shown in Fig.\,\ref{fig4} for the respective locked input angles.

 The values obtained for $\sqrt\kappa$ are shown in the top panel. To compare them with a theoretical estimate, we calculate its value starting from (\ref{eq:kappa}). The torque $\tau$ is caused by the dipole dipole interaction and reads
\begin{equation}
    \vec{\tau} = \vec{m}_2 \times \vec{B}_1.
\label{eq:tau}
\end{equation}
Here $\vec{B}_1$ is the magnetic flux density of the input magnet
\begin{equation}
    \vec{B}_1=\frac{\mu_0}{4\pi}\frac{3\left(\vec{m}_1\cdot\vec{r}\right)\vec{r}-\vec{m}_1 r^2}{r^5},
\label{eq:Bfield}
\end{equation}
while the magnetic moment of the input magnet $\vec{m}_1$, of the output magnet $\vec{m}_2$ and the distance vector $\vec{r}$ are 
\begin{equation}
\begin{split}
    \vec{m}_1&=m_1\begin{pmatrix}
                    -\cos{\alpha}\sin{\Theta} \\
                    \cos{\alpha}\cos{\Theta} \\
                    \sin{\alpha} 
                \end{pmatrix},\\
                \vec{m}_2&=m_2\begin{pmatrix}
                    -\cos{\beta}\sin{\Phi} \\
                    \cos{\beta}\cos{\Phi} \\
                    \sin{\beta} 
                \end{pmatrix},\,
                \vec{r}=r\begin{pmatrix}
                    1 \\
                    0 \\
                    0 
                \end{pmatrix}.
\end{split}
\label{eq:vectors}
\end{equation}

Since the output can only rotate around the shaft axis, only the component $\tau_{\text{ax}}$ of the torque along the shaft is of interest. We calculate it as
\begin{equation}
    \tau_{\text{ax}}=\vec{\tau}\begin{pmatrix}
                    \cos{\Phi} \\
                    \sin{\Phi} \\
                    0 
                \end{pmatrix}.
\end{equation}
With (\ref{eq:Delta}),\,(\ref{eq:tau}),\,(\ref{eq:Bfield}),\,(\ref{eq:vectors}) this becomes
\begin{equation}
    \tau_{\text{ax}} = \frac{\mu_0}{4\pi}\frac{m_1 m_2}{r^3}\left(-\sin{\alpha}\cos{\beta}+\Delta\cos{\alpha}\sin{\beta}\right).
\label{eq:tauax}
\end{equation}

The moment of inertia of the output shaft $I_{\text{ax}}$ is difficult to estimate. Reasons for this are the the spacers on both shaft ends, the additional moment of inertia from the bearing balls, and the different densities between the permanent magnet and the rest of the shaft material. We therefore treat it as a fit parameter in the calculation of $\kappa$. With (\ref{eq:tauax}) we get
\begin{equation}
    \kappa=\frac{\mu_0 m_1 m_2}{4\pi r^3 I_{\text{ax}} }\left(\sin{\alpha_0}\sin{\beta_0}+\Delta\cos{\alpha_0}\cos{\beta_0}\right)
\label{eq:kappaex}
\end{equation}

We measured the magnetic dipole moments of the input and output in a former work as $m_1$\,=\,$m_2$\,=\,$(3.51$\,$\pm$\,$0.11)\,\text{JT}^{-1}$ \cite{Hartung2018}. The distance between the magnets is $r$\,=\,$(40.0$\,$\pm$\,$0.5)\,\text{mm}$. With this, we fit $\sqrt{\kappa}$ via $I_{\text{ax}}$ according to (\ref{eq:kappaex}) to the data for different $\alpha_0$. The result is seen as the red line in the top panel of Fig.\,\ref{fig4}. The red shaded area marks the uncertainty of the fitted curve that stems from the variation of the constants within their measurement accuracy. The fitted value for the moment of inertia is
\begin{equation}
    I_{\text{ax}}=6.45\cdot10^{-5}\,\text{kg}\text{m}^2.
\label{eq:inertia_fit}
\end{equation}
The complete output shaft has a mass of $m$\,=\,$(320$\,$\pm$\,$1)\,\text{g}$, which corresponds to a radius of gyration of
\begin{equation}
    r_\text{gyr}=\sqrt{\frac{I_{\text{ax}}}{m}}=14.2\,\text{mm}.
\label{eq:rad_gyr}
\end{equation}
By approximating the output shaft as a circular cylinder and its mass to be homogeneous, its effective diameter is
\begin{equation}
    d_\text{eff}=2 \frac{r_\text{gyr}}{\sqrt{2}}=40.16\,\text{mm}.
\label{eq:d_eff}
\end{equation}
This is a reasonable result, since the inside diameter of the bearings that hold the shaft is $d_\text{cyl}$\,=\,$(30.0$\,$\pm$\,$0.1)\,\text{mm}$ and the diameter of the spacers on the shaft ends is $d_\text{spacer}$\,=\,$(50.0$\,$\pm$\,$0.1)\,\text{mm}$.

We see in Fig.\,\ref{fig4} that the experimentally obtained values for $\kappa$ are in good agreement with the theoretical prediction. Nevertheless, the fact that the red curve leads to systematically higher values at the maxima and lower values at the minima indicates that the actual shaft angles might be slightly smaller than what we measured at the spacer position.

The middle and bottom panels of Fig.\,\ref{fig4} show the fitted values for $\zeta$ and $\eta$. Their relative variation is up to approximately 100\%. Any dependence on $\alpha_0$ however, we interpret as shortcomings of our setup that hold no further information. We therefore simply take the arithmetic means of both data sets, which are shown as the horizontal red line in the respective graph, and get
\begin{equation}
\begin{split}
    \zeta &= (1.1 \pm 0.4)\,\text{s}^{-1}\\
    \eta &= (23.1 \pm 9.9)\,\text{s}^{-2}.
\end{split}
\label{eq:zeta}    
\end{equation}

The sliding friction coefficient $\mu$ is determined from (\ref{eq:zeta}) and (\ref{eq:inertia_fit}) and yields
\begin{equation}
    \mu=\frac{\eta\,I}{\,F\,\frac{d_\text{cyl}}{2}}=\frac{\eta\,d_{\text{eff}}^2}{4\,g\,d_{\text{cyl}}}=0.031\pm0.014,
\label{eq:mu}
\end{equation}
with $g$ being the gravitational constant. 
For the dry sliding friction coefficient between two surfaces of silicon nitride $(\text{Si}_3\text{N}_4)$, the material of the bearing balls and cage, we find a way higher value of $\mu$\,=\,$0.17$ in literature \cite{bhushan1982silicon}\,\cite{wang2000rolling}\,\cite{dalal1975evaluation}. However, one has to keep in mind that we calculated the friction coefficient for a system with rolling balls instead of sliding ones. Aramaki et al. investigated the friction of a $\text{Si}_3\text{N}_4$ bearing and found that values for $\mu$ vary from 0.01 to 0.05 depending on the applied load and the spinning velocity of the balls \cite{aramaki1988performance}. This is in good agreement with our result.

\section{Output response for finite driving frequencies}
We now want to analyse the response of the output magnet while the input magnet is driven with a constant rotation frequency. From measuring $\beta$ we calculate the angle difference $\delta$ between input and output as
\begin{equation}
    \delta = \beta - \alpha.
\label{eq:delta}
\end{equation}
We do so for parallel rotation axes and different shaft angles $\Theta$\,=\,$\Phi$ and driving frequencies $f$.

\begin{figure}[!t]
\centering
\includegraphics[width=\linewidth]{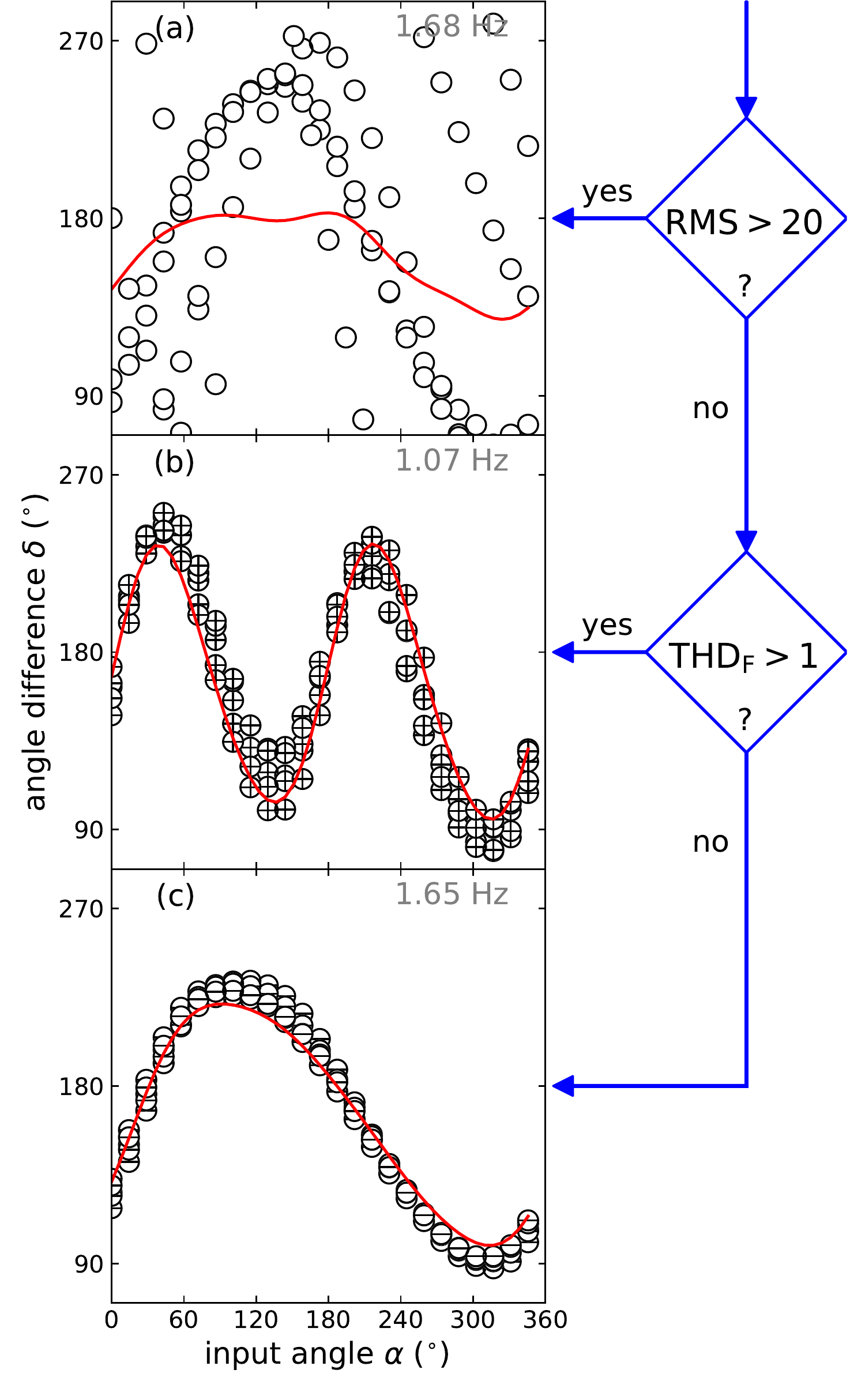}
\caption{The method for differentiating between three types of output rotation. The right-hand side shows a logical flow chart for the decision making. On the left-hand side an exemplary graph for each kind is provided. The $x$-axis shows the rotation $\alpha$ of the input modulo $360^{\circ}$. The $y$-axis shows the angle difference $\delta$. For each plot the response to the last 10 of 100 full rotations of the input is shown. Non-periodic response is marked in (a) by empty circles. Circled crosses in (b) denote $T/2$ periodicity, and circled dashes in (c) $T$ periodicity. The red lines show the results of the harmonic fits (\ref{eq:d_harm}) to the data points.}
\label{fig5}
\end{figure}

Three examples of these measurements are shown for $\Theta$\,=\,$31^{\circ}$ in Fig.\,\ref{fig5}. Each of them features a unique response of the output. In the bottom panel $\delta$ is periodic with a full rotation of the input, which we call $T$-periodic. The middle panel shows $\delta$ to be periodic with a half rotation of the input, i.\,e., $T/2$-periodic. In the top panel the answer of $\delta$ is seemingly non-periodic, chaotic.

We find that after a short settling phase, each of our measurements falls in one of these categories. An objective tool for differentiation is by the method shown in the flowchart on the right-hand side of Fig.\,\ref{fig5}. We first make a multi-harmonic fit $\delta_\text{harm}$ to the data
\begin{equation}
    \delta_\text{harm}=\left(\sum_{i=1}^{7}{a_i\sin{i\,\alpha}+b_i\cos{i\,\alpha}}\right)+b_0
\label{eq:d_harm}
\end{equation}
with $a_{1,...,7}$ and $b_{0,...,7}$ as the fitting parameters. Here $i$\,=\,$1$ marks a response with the fundamental mode of the input while $i$\,=\,$2,...,7$ are the respective higher harmonics. The first decision is done regarding whether the output behaves chaotically or is answering periodically. For this we calculate the root mean square (RMS) of the difference between $\delta$ and $\delta_\text{harm}$ for the last 10 rotations of a measurement where 50 positions of $\alpha$ are detected for each rotation, namely
\begin{equation}
\text{RMS}=\sqrt{\frac{1}{500}\sum_{m=4501}^{5000}{(\delta_m-\delta_\text{harm}(\alpha_i))^2}}.
\label{eq:rms}
\end{equation}

We find that $\text{RMS}>20$ serves well to be characteristic for a chaotic output. To differentiate between $T$ periodicity and $T/2$ periodicity, we calculate the total harmonic distortion $(\text{THD}_\text{F})$ of $\delta_\text{harm}$ as
\begin{equation}
\text{THD}_\text{F}=\frac{\sqrt{\sum_{i=2}^7{(a_i^2+b_i^2)}}}{a_1^2+b_1^2}.
\label{eq:thd}
\end{equation}
The benefits of this definition of the total harmonic distortion were shown by Shmilovitz \cite{shmilovitz2005definition}. For $\text{THD}_\text{F}$\,=\,$1$, the amplitudes of all harmonic modes of $\delta$ together are just as big as the fundamental one. We use $\text{THD}_\text{F}$\,$>$\,$1$ to identify the case of $T/2$ periodicity. For smaller values the fundamental mode prevails and $\delta$ is seen as $T$-periodic.

%

\begin{figure}[!t]
\centering
\includegraphics[width=0.612\linewidth]{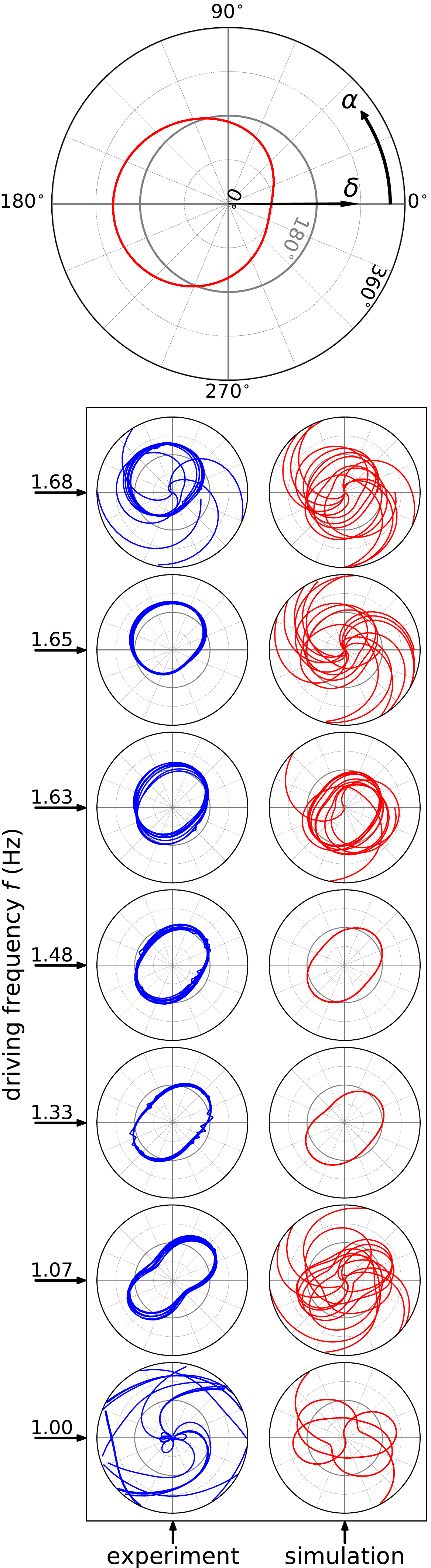}
\caption{Polar diagrams of the angle differences $\delta$ for different driving frequencies of the input $f$, and $\Theta$\,=\,$31^{\circ}$. The graphs each show the last 10 of 100 full rotations of input angle $\alpha$ modulo $360^{\circ}$. On the top an expanded exemplary diagram is shown that contains the simulated curve for 2,08\,Hz. The radial distance is given by $\delta$ and the azimuthal angle by $\alpha$. The arrows on the left-hand side mark the respective input driving frequencies. Experimental results are shown in blue, the simulation in red.}
\label{fig6}
\end{figure}

To further investigate the transition between these phases, $\delta$ is shown in Fig.\,\ref{fig6} for different driving frequencies of the input $f$ at constant shaft angles $\Theta$\,=\,$\Phi$\,=\,$31^{\circ}$. This way of presenting the data is adopted from the work of Borgers et al.\,\cite{borgers2018}. It is explained by a more detailed diagram in the top part of  Fig.\,\ref{fig6}. 

The frequencies shown in the bottom part are chosen as such that the polar diagrams give a good representation of the phases observed in the experiment and displayed on the left-hand side. For each plot the measurement started from the resting position of the input at $\alpha$\,=\,$90^\circ$ and the output at $\beta$\,=\,$270^\circ$. This is followed by a settling phase of 90 input rotations. The data that are shown stem from the next 10 input rotations.

For $f$\,=\,$1.00\,\text{Hz}$, the experimentally observed $\delta$ seems to behave chaotically. When  $f$ is increased up to 1,07\,Hz, $T/2$ periodicity of $\delta$ is observed. The thickness of the line indicates the experimental noise. Both at 1.33\,Hz and at 1.48\,Hz  $T/2$ periodicity can be seen as well --- the pattern clearly is symmetric with a $180^{\circ}$ turn of $\alpha$. This is clearly not the case anymore for 1.63\,Hz and 1.65\,Hz where we enter the $T$-periodic regime. At 1.68\,Hz we observe seemingly chaotic behavior again. 

Simulated results for $\delta$ are shown on the right-hand side for the respective driving frequencies. They were derived using a model similar to (\ref{eq:model}) except for a change in the last summand, because the magnetic torque is now time-dependent. With the use of (\ref{eq:tauax}), (\ref{eq:inertia_fit}) and (\ref{eq:zeta}) we get
\begin{equation}
    \ddot{\beta}+\zeta\dot{\beta}+\eta\,\mathrm{sgn}(\dot{\beta})+g_{\tau}\left(-\sin{\alpha}\cos{\beta}+\Delta\cos{\alpha}\sin{\beta}\right)=0,
\label{eq:modeldyn}    
\end{equation}
with
\begin{equation}
    g_{\tau}=\frac{\mu_0}{4\pi}\frac{m_1 m_2}{I_\text{ax}\,r^3}.
\label{eq:g}    
\end{equation}
We solve this equation for $\beta$ numerically, with discreet time steps
\begin{equation}
    \Delta t = \frac{1}{n\,f},
\label{eq:deltat}
\end{equation}
with $f$ being the driving frequency of the input and $n=50$. 

For 1.00 Hz the simulation differs qualitatively from the experiment: the response is $3\,T$-periodic. At 1.07 Hz the simulation creates a chaotic response. We see a $T$-periodic response at 1.33\,Hz and a $T/2$-periodic response at 14.8\,Hz. For 1,63\,Hz, 1,65\,Hz and 1,68\,Hz the simulated data appears to be chaotic.

In a recent work of Haugen and Edwards it was shown that the free oscillation of two magnetic dipoles in a plane does not feature chaotic behavior, contradictory to their own intuition \,\cite{haugen2020dynamics}. This is not in contradiction with our findings here, because they did not take an externally driven magnet into account. This adds one degree of freedom to the system, thus allowing for chaotic motions. 

In summary, both experiment and simulation create similar dynamical scenarios, but a quantitative match can not be achieved due to the simplicity of our model. 

\begin{figure}[!t]
\centering
\includegraphics[width=\linewidth]{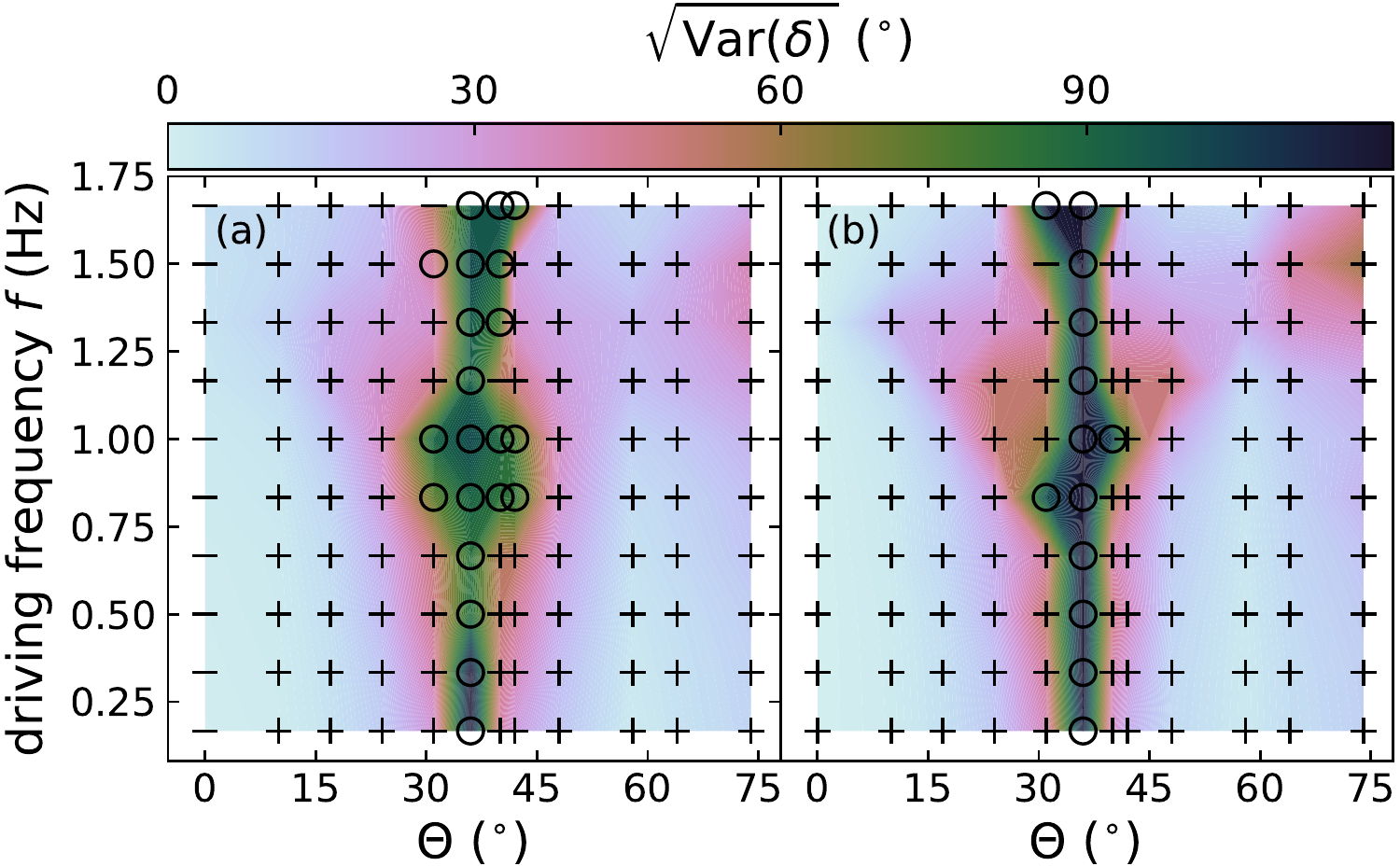}
\caption{Color map of the standard deviation of the angle difference $\delta$ for different shaft angles $\Theta$ and different driving frequencies of the input $f$. The graph (a) on the left shows experimental data. The graph (b) stems from a simulation of the system according to (\ref{eq:modeldyn}) using the parameters of (\ref{eq:zeta}). Each data point is represented by a marker: circles indicate chaotic response of the output, crosses $T/2$ periodicity, and dashes $T$ periodicity.}
\label{fig7}
\end{figure}

In order to further investigate qualitative changes in the dynamics of the output magnet in our experiment, the measurement of $\delta$ is shown in Fig.\,\ref{fig7} as a color map for different parallel shaft angles $\Theta$\,=\,$\Phi$ and driving frequencies $f$. The experimental data shown on the left-hand side can be compared  to a simulation of (\ref{eq:modeldyn}) with appropriate parameters shown on the right-hand side. In both cases, phases of chaotic behavior are surrounded by periodic phases with high standard deviation of $\delta$ which stem from oscillations with high amplitudes.

The shaft angles near $\Theta$\,=\,$0^{\circ}$ and $\Theta$\,=\,$54.74^{\circ}$ --- the cogging-free cases \cite{Schoenke2015,borgers2018} --- are characterised by a minimum of $\text{RMS}_{\delta}$\,=\,$\sqrt{\text{Var}(\delta)}$. Once the shaft angles approach $\Theta$\,=\,$35.26^{\circ}$, we reach a maximum of this value. This is the position where the highest cogging occurs, accompanied by a change of the sense of rotation of output angle $\beta$. We conclude that the cogging of this magnetic gear is an important factor for the onset of chaotic motion.

It is interesting to note that the red highlighted area of higher oscillation amplitudes is widest spread at driving frequencies near $f$\,=\,1.3\,Hz. A possible reason for this could be the reaching of a resonance frequency of the output shaft. This might also explain why chaos is predominantly observed near that frequency.

\begin{figure}[!t]
\centering
\includegraphics[width=\linewidth]{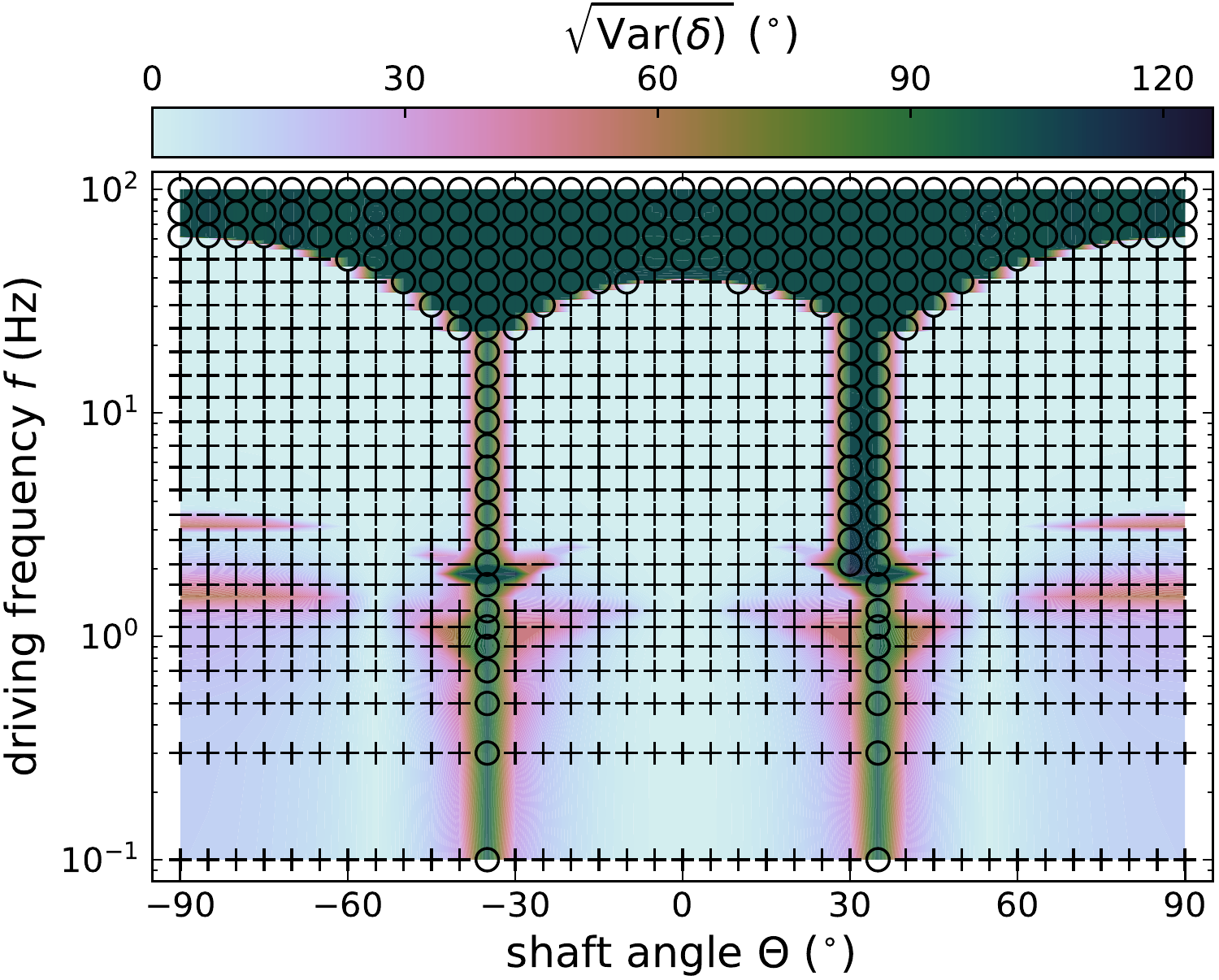}
\caption{Color map of the standard deviation of the angle difference $\delta$ for different shaft angles $\Theta$ and increasing driving frequencies of the input $f$. Only simulated data according to (\ref{eq:modeldyn}) are shown with the use of the system parameters from (\ref{eq:zeta}). The step width in $\Theta$ is $5^{\circ}$, and 0.02\,Hz in $f$. Circles indicate chaotic response of the output, crosses $T/2$ periodicity, and dashes $T$ periodicity.}
\label{fig8}
\end{figure}

While the experimental results summarized in Fig.\,\ref{fig7} are restricted to the parameter range accessible in our experiment, we provide an expanded range in Fig.\,\ref{fig8}, where the data are based on purely numerical simulation. The whole range of shaft angles for parallel alignment is analyzed with increasing driving frequencies for each configuration that range from 0.1\,Hz up to 100 Hz and is shown on a logarithmic scale.

While the symbols and color map are the same as in Fig.\,\ref{fig7}, we now free ourselves from the restriction of using the steady state as an initial condition for each new driving frequency as done in the experiments. Instead, the driving frequency is now increased in a semi-static way for each shaft angle. After each frequency increase of 0.2\,Hz, a waiting time of 1000 rotations is implemented. The starting geometry for each new $\Theta$ is $\alpha$\,=\,$0^{\circ}$ and $\beta$\,=\,$180^{\circ}$. 

A striking feature is the symmetry breaking between $\Theta$\,=\,$-30^{\circ}$ and $\Theta$\,=\,$30^{\circ}$ above $f$\,=\,$20\,\text{Hz}$. This can be explained by our starting configuration, which breaks the mirror symmetry between positive and negative $\Theta$ values. Moreover, it clearly indicates that multi-stability is present in this regime. 

\begin{figure}[!t]
\centering
\includegraphics[width=\linewidth]{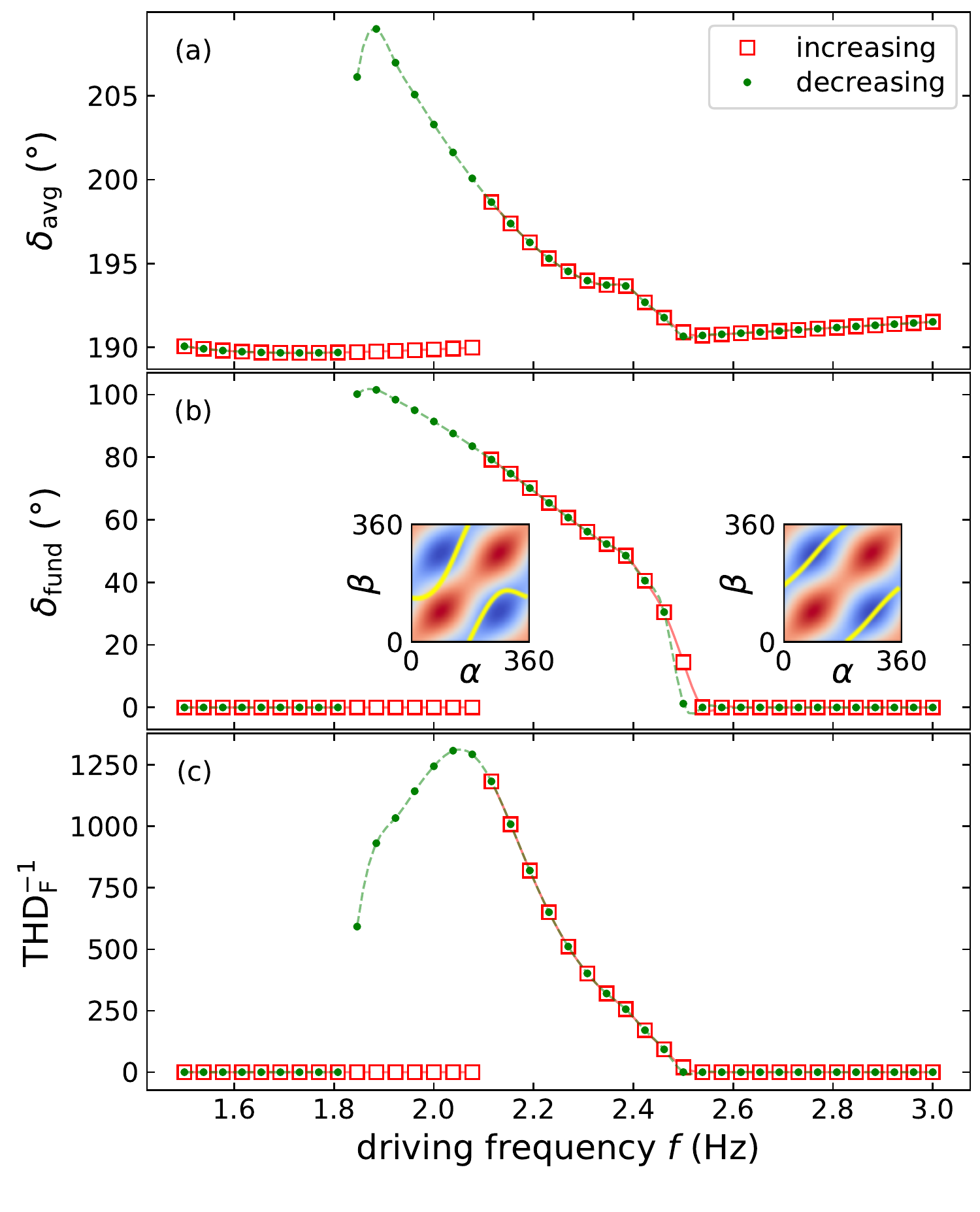}
\caption{The transition between $T$-periodic output response and $T/2$ periodicity according to numerical simulation for $\Theta$\,=\,$25^{\circ}$. The open squares correspond to increasing, the solid circles to decreasing values of the driving frequency. The red solid lines and the green dashed lines are guides to the eye. The left inset shows an example of the phase trajectory from the $T$-periodic regime at $f$\,=\,2.1\,Hz, the right inset of the $T/2$-periodic regime at $f$\,=\,2.7\,Hz. The background color of the insets indicates the strength of the magnetic field energy. The valleys are shown in blue, the hills in red.}
\label{fig9}
\end{figure}

Outside the immediate surrounding of the cogging-free shaft angle configurations, e.\,g.\, at $\Theta$\,=\,$25^{\circ}$, a maximum of $\text{RMS}_{\delta}$ is seen near $f$\,=\,$1.25\,\text{Hz}$. For further increase in driving frequency, the amplitude decreases but forms a second maximum at $f$\,$\approx$\,$2.5\,\text{Hz}$. This marks a window of $T$-periodic response of the output in an otherwise $T/2$-periodic surrounding and can be seen in detail in Fig.\,\ref{fig9}.

\begin{figure}[!t]
\centering
\includegraphics[width=\linewidth]{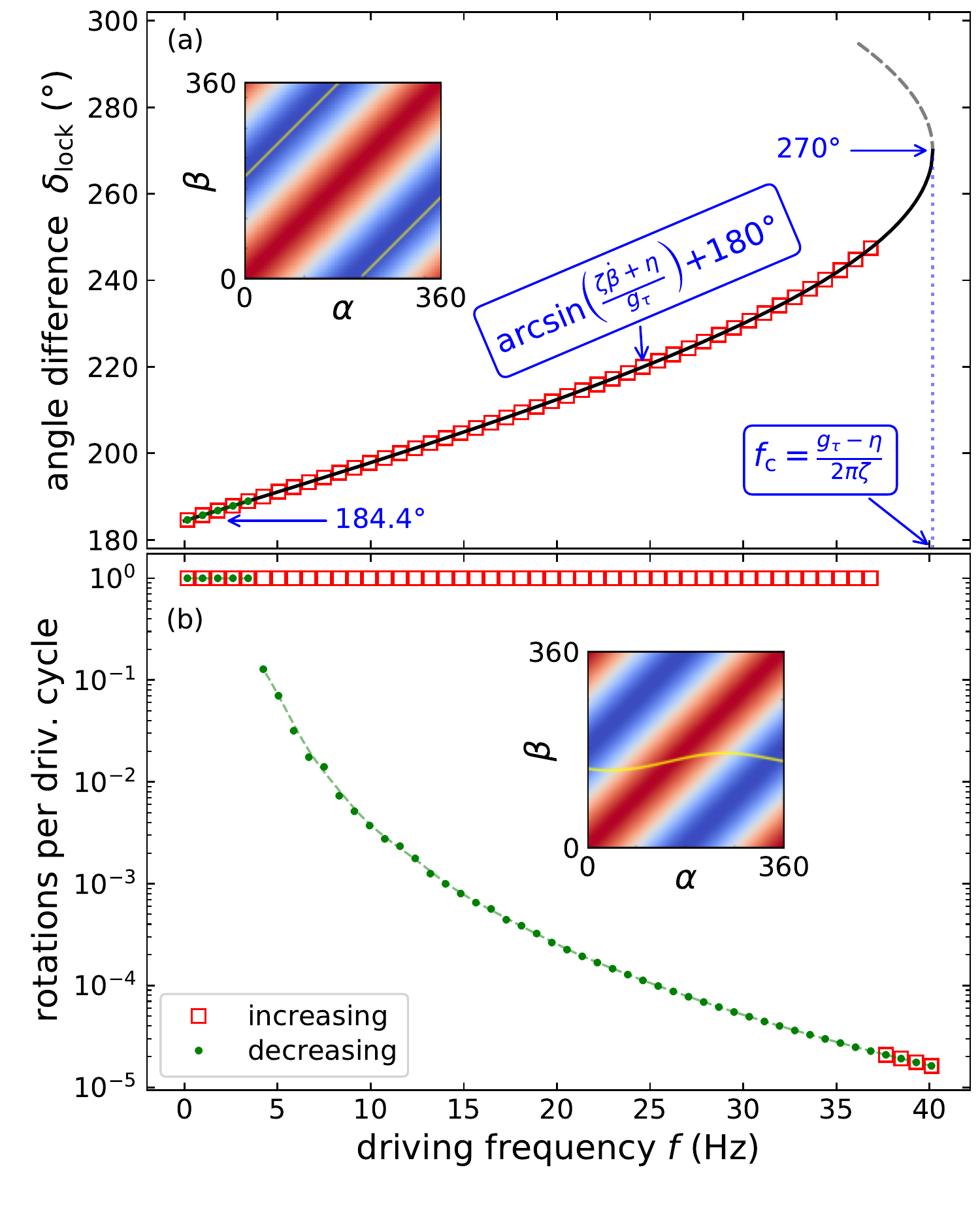}
\caption{The transition between the locked and unlocked operation mode of the gear according to the numerical simulation. The open squares correspond to increasing, the solid circles to decreasing values of the driving frequency. The solid black line in (a) is the stable branch of the analytical solution of the locked state, its unstable branch is indicated as a dashed grey line. The dashed green line in (b) should only serve to guide the eye. The inset in (a) shows an example of the phase trajectory in the locked state at a driving frequency of 6\,Hz, the inset in (b) of the slip-through state at the same driving frequency. The background color of the insets indicates the strength of the magnetic field energy. In the locked state, the yellow trajectory remains in the valley of the minimal energy configuration (blue).  In the slip-through state, the trajectory crosses the hill indicated in red.}
\label{fig10}
\end{figure}

If $f$ is increased even further, $\text{RMS}_{\delta}$ remains small until a certain frequency threshold. Beyond that, the input is no longer locked to the output, the output rather slips through. A more detailed investigation of this transition is shown in Fig.\,\ref{fig10}.

An interesting feature of Fig.\,\ref{fig8} is the transition from $T/2$ periodicity to $T$ periodicity, occurring e.\,g.\ at $\Theta$\,=\,$25\, ^{\circ}$ near $f$\,=\,$2.2\,\mathrm{Hz}$. This period-doubling transition is examined in detail in Fig.\,\ref{fig9}. We simulated this transition by changing the driving frequency in a quasi-static manner, i.\,e., after every frequency change the simulation of (\ref{eq:modeldyn}) was allowed to relax into an equilibrium situation. This protocol was applied both for increasing and decreasing frequency steps to cope with the hysteresis in this transition. 

In the bottom panel of Fig.\,\ref{fig9} the reciprocal of the total harmonic distortion $\text{THD}_{\text{F}}^{-1}$ is shown. Its values are effectively zero below 1.8\,Hz and above 2.5\,Hz. In between these frequencies we see a drastic change towards finite values of $\text{THD}_{\text{F}}^{-1}$ that reach a maximum around 2.1\,Hz. Hysteresis is clearly present between 1.9\,Hz and 2.5\,Hz between increasing and decreasing driving frequencies. At 2.5\,Hz as small deviation between the increasing and decreasing branch can also be observed. This is presumably caused by the critical slowing down of the dynamics near this period-doubling bifurcation.

The middle panel shows $\delta_{\text{fund}}$, the fundamental mode of the discrete Fourier transform of the angle difference $\delta(t)$. Its values are zero where $\text{THD}_{\text{F}}^{-1}$ is zero, and they are finite in the same regime as well. However, for decreasing driving frequencies, $\delta_{\text{fund}}$ reaches its maximum at a lower $f$ than  $\text{THD}_{\text{F}}^{-1}$ at about 1.9\,Hz. This plot is especially well-suited to illustrate that we are dealing with a supercritical period-doubling bifurcation at 2.5\,Hz with the characteristic square root increase of the order parameter $\delta_{\text{fund}}$. The bifurcation at 2.1\,Hz is subcritical, on the other hand. Its unstable branch gains stability in a saddle-node bifurcation at 1.85\,Hz \cite{strogatz1994nonlinear}.

The two insets in the middle panel show the trajectories of $\beta(\alpha)$ and the corresponding magnetic field energy
\begin{equation}
   E(\alpha,\beta)=\frac{\mu_0}{4\pi}\frac{m_1 m_2}{r^3}\left( \sin{\alpha} \sin{\beta}+\Delta \cos{\alpha}\cos{\beta} \right)
\label{eq:field_energy}
\end{equation}
 of the output dipole $\vec{m}_2$ in the field of the input dipole $\vec{m}_1$, color coded in the background. The left inset was taken for decreasing frequencies at $f$\,=\,2.1\,Hz. The evolution in the range $0^{\circ}-180^{\circ}$ of $\beta(\alpha)$ is substantially different from the evolution in the range $180^{\circ}-360^{\circ}$, a clear manifestation of $T$ periodicity.

The right inset shows the behavior for decreasing frequency steps at $f$=2.7\,Hz. The trajectory is now similar in the first and the second half of the driving cycle, a clear indication of $T/2$ periodicity.

The top panel shows the calculated value $\delta_{\text{avg}}$ averaged over one full rotation of the input. It measures the average lag between the output and the input angle. The hysteresis also finds a manifestation in this lag. $\delta_{\text{avg}}$ slightly increases with an increase in driving frequency for $f$\,$>$\,$2.5$\,Hz. This is due to the increased damping caused by the liquid-like friction. Below 1.8\,Hz, one can observe a local minimum of $\delta_{\text{avg}}$. We cannot provide a simple explanation for this minimum, but it might be connected to the resonance phenomenon near 1.25\,Hz, mentioned in the discussion of Fig.\,\ref{fig8}.

Fig.\,\ref{fig8} clearly indicates that the output of the gear might slip through for any $\Theta$, provided that the driving frequency is large enough. Understanding the nature of this transition is of greatest technological interest for the practical application of the gear. Thus we simulated this transition by applying the same procedure used to calculate Fig.\,\ref{fig9}. The results for $\Theta$\,=\,$0^{\circ}$, the fundamental cogging-free coupling geometry, are presented in Fig.\,\ref{fig10}. Starting at a small driving frequency of 1\,Hz, the output is locked to the input in the sense that a constant angle difference $\delta$ is asymptotically achieved. This locked state is illustrated by the yellow line in the upper inset of  Fig.\,\ref{fig10}, which yields from a simulation at a driving frequency of 6\,Hz. The background color of the inset indicates the strength of the magnetic field energy given by (\ref{eq:field_energy}). In the locked state, the yellow trajectory remains in the valley of the minimal energy configuration indicated in blue. Increasing the driving frequency leads to an increase of the locked angle according to
\begin{equation}
    \delta_{\mathrm{lock}} = \mathrm{arcsin} \left(\frac{\zeta\dot\beta+\eta}{g_{\tau}}\right) + 180^{\circ},
\label{eq:delta_lock}
\end{equation}
which is determined from (\ref{eq:modeldyn}) by assuming $\ddot \beta = 0$.

The locked branch starts at a finite value of $184.4^\circ$ determined by the solid state friction for a driving frequency $f\rightarrow 0\,\mathrm{Hz}$.  The branch terminates in a saddle node bifurcation a value of $270^\circ$ for a driving frequency $f_\mathrm{c}$\,=\,$\frac{g_{\tau}-\eta}{2 \pi \zeta}$ determined by both friction coefficients. It is interesting to note that the numerical simulation of this branch loses stability slightly before reaching the saddle node bifurcation located at $270^{\circ}$. This can be explained by the distortion caused by the finite frequency steps of about 1\,Hz in the numerical simulation. 

In the unlocked slip-through state, which is reached after a transient following the instability of the locked state, the output almost ceases to move expect for a relatively small wiggling. The lower inset provides an example for the phase trajectory at a driving frequency of 6\,Hz. At this frequency, a clear back and forth movement of the output can be seen, while the net rotation frequency determined by the difference $\beta(360^\circ)$\,$-$\,$\beta(0^\circ)$ is less than 1\,\% of the driving frequency and thus barely visible. This net rotation frequency is indicated by the solid green circles in Fig.\,\ref{fig10}\,(b). Decreasing the driving frequency leads to an increase of this slow rotation. The slip through branch terminates near 5\,Hz, i.e., the width of the hysteresis spans over $80\,\%$ of the width of the locked state. For most technical applications this hysteresis would presumably have to be avoided. This is done by working below the the saddle-node frequency, namely 5\,Hz in our case. This frequency could be increased by using a stronger magnetic coupling $g_{\tau}$.   

\section{Conclusion and outlook}
In this work, the dynamic response of a particular magnetic gear based on pure dipole-dipole coupling is analysed experimentally for the first time. A mathematical model for the dynamics of the output shaft is proposed, which includes two types of friction in the bearing. This simplified model describes the experimental findings on a semi-quantitative level. In particular, it can reproduce $T$ periodicity, $T/2$ periodicity, and chaotic responses. Moreover, it enables us to understand the nature of the bifurcations between these different states. Most importantly, it helps us to clarify the parameter range for a safe operation of this magnetic gear: The driving frequency needs to be sufficiently low, and shaft angles need to be sufficiently far from $\Theta$\,=\,$35.26^{\circ}$, the angle where the output rotation changes its sign.

The mathematical model revealed $3T$ periodicity in a small parameter range, which has not been seen in the experiment. Understanding this discrepancy is the goal of ongoing work.

\appendices

\section*{Acknowledgment}
It is a pleasure to thank R.~Richter, W.~Sch\"opf, and S.~V\"olkel for valuable hints and discussions. This work has been supported by the Deutsche Forschungsgemeinschaft (DFG) - Project No. 389197450.

\ifCLASSOPTIONcaptionsoff
 \newpage
\fi



\bibliographystyle{IEEEtran}
\bibliography{IEEEabrv,Hartung2020}
%

%
\begin{IEEEbiography}[{\includegraphics[width=1in,height=1.25in,clip,keepaspectratio]{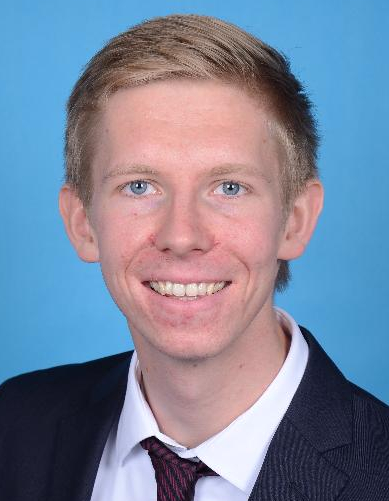}}]{Stefan Hartung}
received the B.S. and M.S degrees in physics from the University of Bayreuth, Germany, in 2014 and 2017 respectively.

Starting in 2017, he has been a PhD student in the department Experimental Physics V at the University of Bayreuth. His research focuses on the interaction of permanent magnets and their ability to form cogging-free gears, the investigation of magnetic dotriacontapoles, and the graphical rectification of magnetisation curves. He also worked on the movement of permanent magnets on a ferrofluid film, and the holographic generation of surface relief gratings.
\end{IEEEbiography}

\begin{IEEEbiography}[{\includegraphics[width=1in,height=1.25in,clip,keepaspectratio]{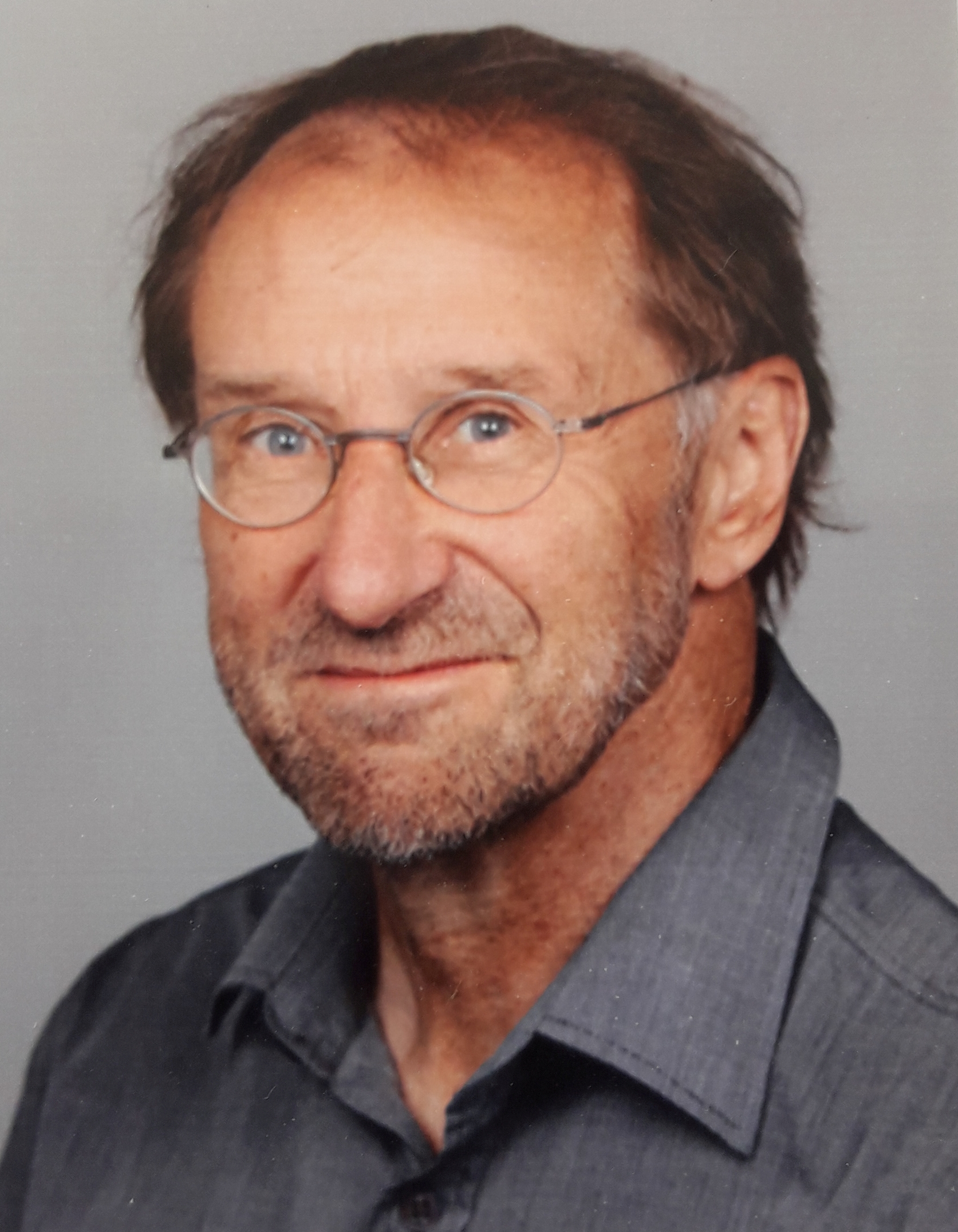}}]{Ingo Rehberg}
received the doctorate in physics from the University of Kiel, Germany, in 1983.

He is currently professor of experimental physics at the University of Bayreuth, Germany. His research focuses on nonlinear dynamics and pattern formation in complex continua like liquid crystals, granular matter and magnetic fluids. The latter ones triggered his interested in dipole-dipole interaction in general, and cogging-free magnetic gears in particular.  
\end{IEEEbiography}






\end{document}